\documentclass[sn-mathphys,Numbered]{sn-jnl}% Math and Physical Sciences Reference Style

\usepackage{graphicx}%
\usepackage{multirow}%
\usepackage{amsmath,amssymb,amsfonts}%
\usepackage{amsthm}%
\usepackage{mathrsfs}%
\usepackage[title]{appendix}%
\usepackage{xcolor}%
\usepackage{textcomp}%
\usepackage{manyfoot}%
\usepackage{booktabs}%
\usepackage{algorithm}%
\usepackage{algorithmicx}%
\usepackage{algpseudocode}%
\usepackage{listings}%
\usepackage{amsmath,bm}
\usepackage{bm}

\theoremstyle{thmstyleone}%
\theoremstyle{thmstyletwo}%

\theoremstyle{thmstylethree}%

\begin{document}

\title[Twisted-photons Distribution at the Axial Channeling]{Twisted-photons Distribution Emitted by Relativistic Electrons at the Axial Channeling}


%\author*[1]{\fnm{First} \sur{Author}}\email{iauthor@gmail.com}


\author*[1]{\fnm{Korotchenko} \sur{K.B.}}\email{korotchenko@tpu.ru}

\author[2]{\fnm{Kunashenko} \sur{Y.P.}}\email{kunashenko@tpu.ru}
\equalcont{These author contributed equally to this work.}

\affil[1]{\orgdiv{Division for Experimental Physics}, \orgname{National Research Tomsk Polytechnic University}, \orgaddress{\street{30 Lenina Ave.}, \city{Tomsk}, \postcode{634050}, \country{Russia}}}

\affil[2]{\orgdiv{Department of mathematics}, \orgname{Tomsk State University of Control Systems and Radioelectronics}, \orgaddress{\street{40 Lenina Ave.}, \city{Tomsk}, \postcode{634050}, \country{Russia}}}


\abstract{Within the framework of quantum electrodynamics, a new method for calculating the radiation of a twisted photon emitted at any angle to the particle velocity has been developed. Using this method, the theory of radiation of a twisted photon by an axially channeled electron at an arbitrary angle to the direction of motion was first created. The twisted-photons angular disibution calculated for the first time.}




\maketitle

\section{Introduction}\label{sec1}

 There are various types of electromagnetic waves in nature: plane, cylindrical and spherical waves. In addition, each electromagnetic wave is characterized by it's polarization. For the first time, this was noticed by Pointing in his work \cite{Poynting}. Quantum mechanics states that every wave type corresponds to a photon. A photon has spin angular momentum. In addition to spin momentum, electromagnetic radiation (photon) can have an orbital angular momentum. As a result, the total angular momentum is the sum of the spin and orbital momentum.

 The projection of the spin onto a given axis (Z axis) can take only two values $m_s = \pm 1$, while the projection of the orbital momentum onto this axis can be equal to $m_l = 0, \pm1, \pm 2,...$ in $\hbar$ units. An example of a photon with a given total angular momentum $m = m_s + m_l$ is a spherical photon. According to \cite{Baranova, Vasara, Bazhenov1, Bazhenov2} there is a new type of waves and particles with the given total angular momentum - twisted waves and particles.

 The ability to generate photons that carry angular momentum opens up new approaches to studying of photonuclear reactions and provides new tools in nuclear physics. The twisted radiation has found numerous applications in both classical and quantum condensed matter optics, high energy physics, optics, etc. (see the review \cite{Vieira} and references therein). Another review of theoretical and experimental works on twisted photons completed by 2018 is given in the work of Knyazev and Serbo \cite{Knyazev}. In \cite{Afanasev, Abramochkin, Torres, Andrews}, problems related to twisted photons were also discussed.

 Serbo and co-authors developed the theory of twisted photons within the framework of quantum electrodynamics. Based on this theory, methods for obtaining beams of twisted photons and possible experiments with them have been studied \cite{Serbo,Serbo1,Serbo2}.

 Along with twisted photons, other twisted particles may exist \cite{Knyazev}, possible experiments with such particles were considered in \cite{Serbo3, Serbo4, Serbo5, Ivanov}.

 In \cite{Bogdanov, Bogdanov1, Bog1, Bog2, Bog3, Bog4, Bog5}, the theory of emission of twisted photons by relativistic charged particles in various undulators, and in other conditions was developed. These works are based on the semiclassical approach developed by Bayer and Katkov \cite{Baier}.

 In these papers, the case when a photon is emitted strictly forward was considered. However, the photon can be emitted at any angle relative to the speed of the particle.

 The passage of particles through oriented crystals is accompanied by various physical phenomena. There are coherent bremsstrahlung and coherent pair production, channeling radiation (CR), the creation of an electron-positron pair by a photon in the continuous potential of the axis (plane) of the crystal, spin rotation of charged channeled particles, diffracted channeled radiation, and others. These phenomena are described in detail in a number of monographs and original articles (see for example \cite{Baier, KoKu, Korotchenko5, Kumakh, Mikelian, Baryshev, Bazylev, Akhiezer, Kimball, Uberall, Lasukov, Lasukov1, Lasukov2, Nitta, Kunashenko, Kunashenko1, Nitta1, Kunashenko2, Kunashenko3} and references therein). Special interest is related to the possibility of creation a powerful source of electromagnetic radiation based on the phenomena of coherent bremsstrahlung and channeling radiation \cite{Kumakh, Mikelian, Baryshev, Bazylev, Akhiezer, Kimball, Uberall, Lasukov}.

 Calculation of the radiation spectrum requires knowledge of the wave functions of a channeled electron in a crystal. This wave function is a product of the longitudinal wave function and the transverse one. Due to the periodic arrangement of the crystal axis (planes), the transverse wave function should be Bloch-wave. In our previous works \cite{Korotchenko1, Korotchenko2, Korotchenko3, Korotchenko4, Korotchenko5} we developed a new method for calculating Bloch wave functions for planar and axial crystal orientations, the details of which are given in \cite{Korotchenko6}.

 In this work, the theory of radiation of a twisted photon at an arbitrary direction with respect to the motion on an axially channeled electron is developed for the first time. Our consideration is based on the use of the approach describing twisted photons \cite{Knyazev, Serbo,Serbo1,Serbo2} and the approach \cite{Korotchenko} describing CR.

 The developed theory can be generalized to describe the radiation of other types of twisted photons.

\section{Twisted-photon wave function}\label{sec3}

 CR occurs when electrons are channeled. It is emitted in an arbitrary direction relative to the crystal axis. It is quite natural to assume that a twisted photon can also be emitted in an arbitrary direction.

 The wave function of a twisted photon has been studied in detail by Serbo and coauthors \cite{Knyazev, Afanasev, Serbo, Serbo1}. However, we will begin our consideration with the TW-photon wave function in order to rewrite it in a form convenient for our purpose.

 Below we call the described process ``Twisted radiation during channeling'' and introduce the designation TWcr.
\begin{figure}[h]
\centering\noindent
\includegraphics[width=6cm]{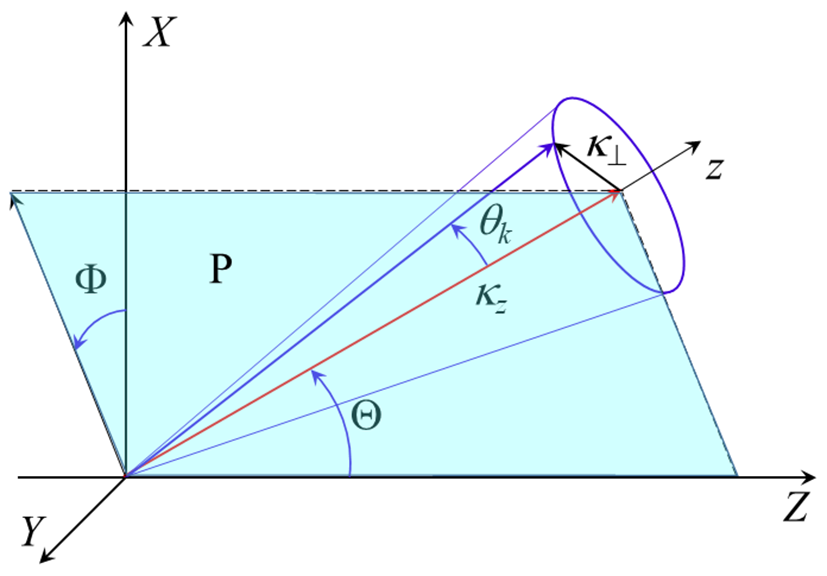}
\caption{The geometry of TWcr emission: where $Z$ is the channeling axis and $z$ is the TWcr-photon axis (directed along the $z$ projection of the photon momentum).}\label{fig:1}
\end{figure}

 In Fig.~\ref{fig:1} polar $\Theta$ and azimuthal $\Phi$ angles determine the directions of the longitudinal momentum $\kappa_z$ of the twisted photon relative to the reference frame associated with the crystal. The $Z$-axis of the coordinate system associated with the crystal is directed along the crystal axis. The coordinate system associated with the twisted photon has the z-axis directed along the vector $\kappa_z$.

 Following the ideas of V.G.Serbo, we chose the wave function of the TW-photon (see, for example \cite{Serbo}) in the form
\begin{equation}
    {\bf A}_{TW} = \int a_{\varkappa m}(\kappa_{\perp}){\bf A}({\bf r},\omega) \frac{d^2\bm{\kappa_{\perp}}}{(2\pi)^2}\: , \label{eq01}
\end{equation}
 with the Fourier amplitude $a_{\varkappa m}(\kappa_{\perp})$
\begin{equation}
 a_{\varkappa m}(\kappa_{\perp}) = \bm{i}^{-m} e^{\bm{i} m \varphi_k} \sqrt{\frac{2\pi}{\kappa_{\perp}}}\delta(\kappa_{\perp} - \varkappa)\: . \label{eq02}
\end{equation}

 Here $\omega$ is the TW-photon frequency, $\kappa_{\perp}$ is the transverse component of the wave vector (see Fig.~\ref{fig:1}).

 If we use the three-dimensional transverse gauge ($\nabla {\bf A}({\bf r},\omega) = 0$) then the plane photon wave function is
\begin{equation}
    {\bf A}({\bf r},\omega) = \sqrt{\frac{2\pi c^2\hbar}{\omega V_{TW}}} e^{\bm{i}(\omega t - {\bm{\kappa}}{\bf r})} {\bm{\varepsilon}}_{\kappa \Lambda}\: , \label{eq03}
\end{equation}
 where ${\bm{\varepsilon}}_{\kappa \Lambda}$ is the polarization vector, $\bm{\kappa}$ is the photon wave vector and $V$ is the normalized volume.

 The polarization vector ${\bf\varepsilon}_{\kappa\Lambda}$ of a twisted photon in the coordinate system associated with the photon (see Fig.~\ref{fig:1}) has the form:
\begin{equation}
    {\bm{\varepsilon}}_{\kappa \Lambda} = \sum_{m_s=-1}^{1}e^{-\bm{i}m_s \varphi_\kappa} d_{m_s\Lambda}^{1}(\theta_\kappa)\bm{\chi}_{m_s}\: , \label{eq04}
\end{equation}
 where $d_{\sigma\Lambda}^{1}(\theta_\kappa)$ is the Wigner (small) d-matrix and ${\bm{\chi}_{m_s}}$ are the eigenvectors of the projection  of the spin operator of a plane-wave photon
\begin{equation}
    {\bm{\chi_0}} = (0,0,1)\: , \:  \: {\bm{\chi}}_{\pm 1} = \mp(1,\pm \bm{i},0)\: . \label{eq05}
\end{equation}

 Substituting (\ref{eq03}, \ref{eq04}) into (\ref{eq01}), we get
\begin{equation}
    {\bf A}_{TW} = \bm{i}^{-m}{\bm{\varepsilon}}_{\kappa\Lambda}\sqrt{\frac{c^2\varkappa\hbar}{\omega V_{TW}}} \int e^{\bm{i}\varkappa r_{\perp}}e^{\bm{i}\kappa_z z} e^{\bm{i} m\varphi_k} \frac{d\varphi_k}{2\pi}\: , \label{eq06}
\end{equation}

 The normalization volume for a twisted photon, according \cite{Serbo, Serbo1}, differs from one for a plane photon and equals $V_{TW} = L_z R/\pi$.

\section{Twisted-photon radiation probability}\label{sec:3}

\begin{figure}[h]
\centering\noindent
\includegraphics[width=7cm]{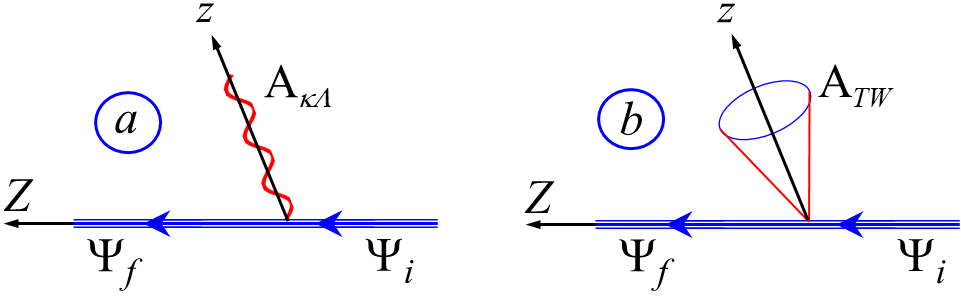}
\caption{a) Feynman diagram CR. b) Feynman diagram TWcr. The $Z$-axis is directed in the direction of the electron velocity, and the $z$-axis is the TWcr photon axis (direction of the $\kappa_z$ projection of the photon momentum).}\label{fig:2}
\end{figure}

 Calculation of the twisted photon radiation probability from a channeled electron is very similar to calculation of CR (see for example \cite{Baier, Kumakh, Baryshev, Bazylev, Akhiezer, Kimball, Korotchenko}  and references therein). Feynman diagrams for CR and TWcr are shown in Fig.~\ref{fig:2}

 The matrix element of the radiation probability is
\begin{equation}
    M_{fi} = -e \int{\bf j}_{fi}{\bf A}_{TW}d^3{\bf r}\: . \label{eq07}
\end{equation}
 Here ${\bf j}_{fi}({\bf r}) = \bar{\psi}_f({\bf r}){\bm{\gamma}}\psi_i({\bf r})$  is the ``current operator'', $\bm{\gamma}$ are the Dirac matrices, $\psi_i$ and $\psi_f$ are the wave functions of the channeled electron at the initial and final states, ${\bf A}_{TW}$ is the TW-photon wave function (\ref{eq06}). Using the matrix element (\ref{eq07}), we  write down the probability of radiation
\begin{equation}
    dW_{fi} = \frac{2\pi}{\hbar}|M_{fi}|^2\delta(\mathcal{E}_i - \mathcal{E}_f - \hbar\omega)
    d\rho\: , \label{eq08}
\end{equation}
 where $d\rho$ is the number of final states. For the twisted photon emission by a channeled electron $d\rho = d\rho_{TW} d\rho_e$, where $d\rho_{TW} = (R d\varkappa/\pi) (L_z d\kappa_z/2\pi)$, with $\hbar\kappa_z$ being the longitudinal momentum, and $\hbar\varkappa$ being the transverse momentum of the twisted photon. The number of final states of the channeled electron is $d\rho_e = (L dp_z)/((2\pi\hbar))$, and $p_z$ is the longitudinal momentum of the electron. Here, we took into account that the longitudinal motion of the channeled electron is free at the final state.

 Using the following relations for the TW-photon
\begin{equation}
    \hbar\omega = c\hbar\sqrt{\varkappa^2 + \kappa_z^2}, \: \: \: \tan\theta_\kappa = \frac{\varkappa}{\kappa_z}\: ,  \label{eq09}
\end{equation}
 it is convenient to rewrite $d\varkappa d\kappa_z$ as $d\varkappa d\kappa_z = \omega d\omega d\theta_\kappa/c^2$. Then we get
\begin{equation}
    dW_{fi} = \frac{2\pi}{\hbar}|M_{fi}|^2\delta(\mathcal{E}_i - \mathcal{E}_f - \hbar\omega) \frac{\omega}{c^2}\frac{d\omega d\theta_\kappa}{2\pi^2}\frac{L dp_z}{2\pi\hbar}
    \: . \label{eq10}
\end{equation}

 After we substitute (\ref{eq06}) into the matrix element (\ref{eq07}) we will face the following problem: the channeled electron wave functions $\psi_{i(f)}(\mathbf{r})$ are written in the coordinate system associated with the crystal, while the wave function ${\bf A}_{TW}$ is written in the coordinate system associated with the TW-photon. Therefore we need to transform the wave vector ${\bf A}_{TW}$ into the crystal coordinate system. To do this, the rotation matrix $m_{\Theta,{\bf X},{\bf Y}}$ can be used
% For this, one can use the rotation matrix
\begin{equation}
 \hspace{-7mm} m_{\Theta,{\bf X},{\bf Y}} = \left(
\begin{tabular}{c c c|c}
$\frac{{\bf {\bf X}}^2\cos\Theta + {\bf Y}^2}{{\bf X}^2+{\bf Y}^2}$ & $\frac{{\bf X}{\bf Y}(\cos\Theta-1)}{{\bf X}^2+{\bf Y}^2}$ & $\frac{{\bf X}\sin\Theta}{\sqrt{{\bf X}^2+{\bf Y}^2}}$ & 0 \\
$\frac{{\bf X}{\bf Y}(\cos\Theta-1)}{{\bf X}^2+{\bf Y}^2}$ & $\frac{{\bf X}^2+{\bf Y}^2\cos\Theta}{{\bf X}^2+{\bf Y}^2}$ & $\frac{{\bf Y}\sin\Theta}{\sqrt{{\bf X}^2+{\bf Y}^2}}$ & 0 \\
$-\frac{{\bf X}\sin\Theta}{\sqrt{{\bf X}^2+{\bf Y}^2}}$ & $-\frac{{\bf Y}\sin\Theta}{\sqrt{{\bf X}^2+{\bf Y}^2}}$ & $\cos\Theta$ & 0 \\
\hline
0 & 0 & 0 & 1 \\
\end{tabular}
\right) . \label{eq11}
\end{equation}

 Acting with this matrix on the matrix ${\bm{\chi_0}}$ and ${\bm{\chi}}_{\pm}$, we have vectors rotated by $\Theta$ radians in the plane spanned by $(0, 0, 1)$ and $(X, Y, 0)$ (the plane $P$ in Fig.~\ref{fig:1} - i.e. ${\bf X} = \sin\Theta\cos\Phi$ and ${\bf Y} = \sin\Theta\sin\Phi$) . The resulting vectors are the ``basic'' vectors for the TWcr-photon in the coordinate system associated with the crystal.

 From our point of view, instead of using vectors $m_{\Theta,{\bf X},{\bf Y}}[{\bm{\chi}}_{\pm}]$, it is more convenient to use the following vectors $e^{\mathbf{-i}m_s\Phi}m_{\Theta,{\bf X},{\bf Y}}[{\bm{\chi}}_{\pm}]$.
 These vectors are rotated by an angle $\Phi$ around the $z$ axis, which means that, they are also photon polarization vectors. The advantage of the new vectors is that they coincide with the polarization vectors of CR photons \cite{Korotchenko}.

 On the other hand, the polarization vector of a CR-photon can be constructed using the Wigner (small) d-matrix \cite{Varsh}
\begin{equation}
   {\bm{\varepsilon}}_{CR}^\Lambda = \sum_{m_s=-1}^{1}e^{-\bm{i} m_s \Phi} d_{m_s\Lambda}^{1}(\Theta){\bm{\chi}}_{m_s}\: . \label{eq12}
\end{equation}

 The ${\bm{\varepsilon}}_{CR}^\Lambda$ vectors and the unit vector in the $z$ direction can be used as the ``basic'' vectors of the TWcr-photon.

 The TWcr-photon polarization vectors in the coordinate system associated with the crystal have the form
\begin{equation}
   {\bm{\varepsilon}}_{\kappa \Lambda}^{TW} = \sum_{m_s=-1}^{1}e^{-\bm{i} m_s \varphi_\kappa} d_{m_s\Lambda}^{1}(\theta_\kappa) {\bm{\varepsilon}}_{CR}^{m_s}\: . \label{eq13}
\end{equation}

 It's simple to verify that the polarization vectors (\ref{eq13}) and the $e^{\mathbf{-i}m_s\Phi}m_{\Theta,{\bf X},{\bf Y}}[{\bm{\chi}}_{\pm}]$ vectors obtained by matrix (\ref{eq11}) coincide.

 The TWcr-photon wave function (in the $XYZ$ crystal coordinate system) is
\begin{equation}
    {\bf A}_{TW}^{cr} = \bm{i}^{-m}\sqrt{\frac{c^2\varkappa\hbar}{\omega V_{TW}}}\int e^{\bm{i} m\varphi_\kappa} e^{\bm{i}\left({\mathbf{K\cdot R}}\right)}\bm{\varepsilon}_{\kappa\Lambda}^{TW}\frac{d\varphi_\kappa}{2\pi}\: , \label{eq14}
\end{equation}
 where $\mathbf{R} = (X,Y,Z)$ and $\mathbf{K} = (\kappa_X,\kappa_Y,\kappa_Z)$ is the wave vector of the TW-photon in the coordinate system $XYZ$. The components of vector $\mathbf{K}$ are equal
\begin{align}
    \kappa_X & = \kappa_z\sin\Theta\cos\Phi + \varkappa(\cos\Theta-1)\cos\Phi\sin\varphi_\kappa\sin\Phi + \nonumber\\
    & \varkappa\cos\varphi_\kappa(\cos\Theta\cos^2\Phi + \sin^2\Phi)\: , \nonumber\\
    \kappa_Y & = \kappa_z\sin\Theta\sin\Phi + \varkappa(\cos\Theta-1)\cos\Phi\cos\varphi_\kappa\sin\Phi + \nonumber\\
    &  \varkappa\sin\varphi_\kappa(\cos\Theta\sin^2\Phi + \cos^2\Phi)\: , \nonumber\\
    \kappa_Z & = \kappa_z\cos\Theta - \varkappa\cos(\varphi_\kappa - \Phi)\sin\Theta\: . \label{eq15}
\end{align}

 The modulus of the TWcr-photon wave vector $\mathbf{K}$ is consistent with the modulus of the vector $|\mathbf{\kappa}|$ in the TW photon coordinate system (see Fig.~\ref{fig:1}))
\begin{equation}
   |(\kappa_X,\kappa_Y,\kappa_Z)| = \sqrt{\varkappa^2 + \kappa_z^2} = \frac{\omega}{c}\: . \label{eq16}
\end{equation}

 Equation (\ref{eq14}) represents the desired form of the TW-photon wave function. This function allows one to calculate the TWcr for arbitrary emission angles.

\section{The calculation of the radiation matrix element}\label{sec:4}

 The wave function of a channeled electron in the quantum state $n$ ($n = i \rightarrow$ initial and $n = f \rightarrow$ final state) can be written in the form \cite{Kumakh}
\begin{align}
 \label{eq17}
 \psi_n(\mathbf{r}) & = \sqrt{\big(E_{n \|} + mc^2\big)\big/E_{n\|}}U\phi_n(\mathbf{r}_{\perp}) e^{\bm{i} p_{n\|}r_{\|}/\hbar}/\sqrt{L}\: , \\
  U & = \left(
\begin{array}{c}
  w \\
  \frac{\bm{\sigma}\cdot\hat{\mathbf{p}}c}{c^2 m + E_{n\|}} w \nonumber
\end{array}
  \right)\: .
\end{align}
 Here $w$ is the 2D-spinor normalized by the condition $w^{+}w = 1$, $\bm{\sigma}$ are the Pauli matrices. The transverse wave function $\phi_i(\mathbf{r}_{\perp})$ describes the quantum states of a relativistic channeled particle and satisfies the Schr\"odinger-like equation \cite{Baryshev, Kimball} with a relativistic mass $\gamma m$ ($\gamma = E_{\|}/m c^2$ is the Lorentz factor).

 Now ``current operator'' ${\bf j}_{fi}$ becomes
\begin{equation}
    {\bf j}_{fi} = C_{fi} \phi_f^{\ast}(\mathbf{r}_{\perp}) U^{+}U \phi_i(\mathbf{r}_{\perp}) e^{\bm{i}\Delta p_{fi\|}r_{\|}/\hbar}/L\: , \label{eq18}
\end{equation}
 where denotated $\Delta p_{fi\|} = p_{i\|} - p_{f\|}$ and
\begin{align}
    U^{+}U & = w^{+}(\bm{\sigma}(\bm{\sigma}\cdot\hat{\mathbf{p}}) + (\bm{\sigma}\cdot\hat{\mathbf{p}})\bm{\sigma})w \nonumber\\
    C_{fi} & = \sqrt{\frac{E_{i\|} + mc^2}{2 E_{f\|}}} \sqrt{\frac{E_{f\|} + mc^2}{2 E_{f\|}}} \: . \label{eq19}
\end{align}
 A simple calculation gives that $U^{+}U = 2\hat{\mathbf{p}}$.

 Taking into account that for relativistic electrons $E_{i\|}\simeq E_{f\|} = E$, for the ``current operator'' (\ref{eq17}) we obtain
\begin{equation}
    {\bf j}_{fi} = \frac{c}{E} \phi_f^{\ast}(\mathbf{r}_{\perp})\hat{\mathbf{p}} \phi_i(\mathbf{r}_{\perp}) e^{\bm{i}\Delta p_{fi\|}r_{\|}/\hbar}/L\: . \label{eq20}
\end{equation}

 Now the matrix element of TWcr-photon emission is equal to
\begin{align}
  M_{fi}^{cr} &= \bm{i}^{-m}\frac{c\hbar}{2}m_{fi}^{cr}\sqrt{\frac{\alpha}{\pi LR}\sin\theta_{\kappa}}\: , \label{eq21}\\
  m_{fi}^{cr} &= \sum_{m_s=-1}^1\int \bm{\alpha}_{fi} d_{m_s\Lambda}^{1}(\theta_\kappa){\bm{\varepsilon}}_{CR}^{m_s} e^{\bm{i}(m-m_s)\varphi_k} e^{\bm{i}\kappa_Z(\varphi_k)Z} d\varphi_k \nonumber\\
  & \frac{1}{L}\int e^{-\bm{i}\frac{\Delta p_{fi,Z}}{\hbar}Z}dZ\: ,\nonumber
\end{align}
 with
\begin{equation}
    {\bm{\alpha}}_{fi} = \frac{c}{E}\int e^{\bm{i}\bm{\kappa_{\perp}} r_{\perp}}\phi_i(r_{\perp})^{\ast}\hat{\mathbf{p}}_{\perp} \phi_f(r_{\perp})dr_{\perp}\: . \label{eq22}
\end{equation}

 Due to the crystal periodicity in the directions perpendicular to the crystal axis, the channeled electron transverse wave function should be Bloch one \cite{Ashcroft, Kittel}. For an axial channeling, it can be represented as \cite{Korotchenko}
\begin{equation}
 \phi_{i,i_n}(x, y) = e^{-\bm{i}\frac{\pi i_n (x+y)}{5 a_p}} \sum_{m_i,n_i}
 e^{-\bm{i}\left(\frac{4\pi m_i x}{a_p} + \frac{4\pi n_i y}{a_p}\right)}C_{i,i_n}^{m_i,n_i}\:, \label{eq23}
\end{equation}
 where $C_{i,i_n}^{m_i,n_i}$ are the Fourier components of the wave function, $m_i$ and $n_i$ are the Fourier components numbers for the $i$-th energy band of the channeled electron transverse motion, $a_p$ is the lattice constant.

 The energy levels of the transverse motion are the energy bands. To take this fact into account, we split each band into $n = 10$ parts. The $i_n$ index corresponds to the $n$-th section of the $i$-th energy band.

 As we pointed out in \cite{Korotchenko}, wave functions (\ref{eq16}) allow precise analytical calculations of CR beyond the dipole approximation. The same is true for TWcr. The huge difference between the energy of the emitted TWcr photon and the longitudinal energy of the electron again allows us to use the dipole approximation.

 Therefore, matrix elements (\ref{eq21}) can be rewritten in the following way \cite{Korotchenko}
\begin{equation}
 {\bm{\alpha}}_{fi} = \Big(\frac{\Omega_{fi}}{c}\kappa_Y(\varphi_\kappa),\: \frac{\Omega_{fi}}{c}\kappa_X(\varphi_\kappa),\: \beta \kappa_X(\varphi_\kappa) \kappa_Y(\varphi_\kappa)\Big) \langle XY\rangle_{fi}\: , \label{eq24}
\end{equation}
 where $\beta = v_{\|}/c$ ($v_{\|}$ is the electron velocity), $\hbar\Omega_ {fi} = \mathcal{E}_i - \mathcal{E}_f$ and $\varepsilon_ {i (f)}$ is the energy of the $i (f)$-th transverse quantum state, the $\langle XY\rangle_{fi}$ equals \cite{Korotchenko}
\begin{equation}
 \langle XY\rangle_{fi} =
  C_{i,i_n}^{m_i,n_i}C_{f,f_n}^{m_f,n_f}\frac{(-1)^{m_f+m_i+n_f+n_i}a_p^2}{16 \pi^2 (m_f-m_i)(n_f-n_i)}\: . \label{eq25}
\end{equation}

 Using equations (\ref{eq23}-\ref{eq24}) for ${\bm{\alpha}}_{fi}$, we find that matrix element (\ref{eq21}) contains integration only over the ``internal'' variables TWcr-photon $\varphi_\kappa$ and $Z$. This is easy to verify if in (\ref{eq07}) integration over the vector $\bm{\kappa_{\perp}}$ is replaced by integration over the polar coordinates $\varkappa$ and $\kappa_{\perp}$ of the vector $\bm{\kappa_{\perp}}$\footnote{According to (\ref{eq02}), the integral over $\bm{\kappa_{\perp}}$ is calculated using the delta function $\delta(\kappa_{\perp} -\varkappa)$.}

 If we were dealing with an ordinary CR, then, in the matrix element (\ref{eq21}) the integral over $\varphi_\kappa$ would be absent and only the integral over $Z$ would remain. It is well known that this integral leads to the law of conservation of longitudinal CR momentum.

 Let us consider for a moment the TWcr-photon as the sum of ordinary photons whose wave vectors lie on the cone of the TWcr photon (see Fig.~\ref{fig:1}). Each photon has its own wave vector. This vector is described by its own angle $\varphi_\kappa$ and, therefore, the photon has its own law of conservation of longitudinal momentum. But the TWcr-photon ``contains'' all such photons. The longitudinal momentum conservation law is not satisfied for all photons simultaneously. Accordingly, for the TWcr-photon itself there is no law of conservation of longitudinal momentum.

 If the angle $\theta_\kappa$ of the TWcr-photon tends to zero, then the cone of vectors that create the TWcr-photon degenerates into a vector directed along the $z$-axis. This means that TW-photon becomes odinary photon. As a result, the dependence on $\varphi_\kappa$ disappears, and the integral over $\varphi_\kappa$ becomes equal to $2\pi$.

 Compared to \cite{Bogdanov, Bogdanov1, Bog1, Bog2} the matrix element $M_{fi}^{cr}$ additionally depends on the angles $\Theta$ and $\Phi$, which describe the direction of the vector $\kappa_z$ (Fig.~\ref{fig:1}). In the papers \cite{Bogdanov, Bogdanov1, Bog1, Bog2}, radiation was considered only strictly forward at the angle $\Theta = 0$. The angles $\Theta$ and $\Phi$ are additional parameters to the usual variables of the wave function of the TWcr-photon. These parameters appear as a result of coordinate transformation (\ref{eq11}, \ref{eq15}).

 Taking into account that $m_o = m-m_s-n \geq 0$, for the integral (\ref{eq21}) over $Z$ we obtain
\begin{align}
  & intJ_{m_o} = \int_0^{\infty} e^{\bm{i} Z(\Delta - A)} J_{m_o}(Z B) dZ = \mathcal{H}[\Delta-A]
  \nonumber\\
 &\frac{\bm{i}}{\sqrt{(\Delta-A)^2 - B^2}}
 \left(\frac{\bm{i} B}{\sqrt{(\Delta-A)^2 - B^2}+\Delta-A}\right)^{m_o} \label{eq26}
\end{align}
 where $A = (\omega/c)\cos\Theta\cos\theta_\kappa$, $B = (\omega/c)\sin\Theta\sin\theta_\kappa$, $\Delta = \Delta p_{fi}/\hbar$ and $\mathcal{H}[...]$ is the Heaviside step function.

 During the calculation of the integral (\ref{eq21}) over $\varphi_\kappa$, we used the following table integrals \cite{Ryzhik}
\begin{align}
 &\int\cos (f + n\varphi_\kappa) e^{-\bm{i}(m-m_s)\varphi_\kappa -A_z \cos(\varphi_\kappa -\Phi)}d\varphi_\kappa = \nonumber\\
 &\Big(e^{2\bm{i}(f + n \Phi)} J_{m-m_s-n}(A_z)+\bm{i}^{2 n}J_{m-m_s+n}(A_z)\Big) \nonumber\\
 &\pi \bm{i}^{m_s-n} e^{-\bm{i}(f + \Phi(m-m_s+n))}\: , \label{eq27}\\
 &\int\sin (f + n\varphi_\kappa) e^{-\bm{i}(m-m_s)\varphi_\kappa -A_z \cos(\varphi_\kappa -\Phi)}d\varphi_\kappa = \nonumber\\
 &\Big(e^{2\bm{i}(f + n \Phi)} J_{m-m_s-n}(A_z)-e^{\bm{i} n\pi}J_{m-m_s+n}(A_z)\Big) \nonumber\\
 &\pi\bm{i}^{1-m_s-n} e^{-\bm{i}(f + \Phi(m-m_s+n))} \: . \label{eq28}
\end{align}
 where $J$ is the Bessel function $A_z = \kappa_{\rho} Z\sin\Theta$, $f$ is a linear function of the angles $\Theta$ and $\Phi$, $n = 0, \pm 1, \pm 2$.

 From the property of Bessel functions it follows $intJ_{-m_o}(\Delta) = (-1)^{m_o} intJ_{m_o}(\Delta)$ \cite{Bateman}. This means that for both positive and negative values $m_o = m-m_s-n$, the square of matrix element $M_{fi}^{cr}$ has the same values.

\section{TWcr-photons distributions}\label{sec:5}

 Let us sum up the probabilities of emission of TWcr-photons by polarization (helicity).

 Similarly \cite{Baier, Kumakh, Baryshev, Bazylev, Akhiezer, Kimball}, we find that the energy conservation law for TWcr is $\hbar\omega \approx \hbar(\beta c\Delta + \Omega_{fi})$.

 From the law of conservation of energy it follows that $\Delta \times intJ_{m_o}$ depends only on the angles $\Theta$ and $\Phi$, i.e. $intJ_{m_o} = f(\Theta,\Phi)/\Delta$.

 After integrating of the probability (\ref{eq20}) over the TWcr-photon frequencies $\omega$ and over the longitudinal momentum of the electron $dp_f$ (i.e. $\Delta$), we get\footnote{The integration over $\Delta$ is carried out in the range from $0$ to $\Delta_{max} = \omega_{max}(\sin\Theta\sin\theta_{\kappa} + \cos\Theta\cos\theta_{\kappa})/c$, where $\omega_{max} = \Omega_{fi}/(1-\beta)$ is the maximum value of the CR-photon frequency.}
\begin{equation}
 \frac{dW_{fi}}{d\theta_{\kappa}} = \frac{\alpha}{4\pi}\sin\theta_{\kappa}(c\beta\Delta_{max}) \|m_{fi}^{cr}(c\beta\Delta_{max})\|^2 P_i(\theta_0)\: .
\label{eq29}
\end{equation}
 Here we took into account the initial population of the transverse energy levels $P_i(\theta_0)$ of the ith energy band of the transverse electron motion ($\theta_0$ is the angle of electron momentum relative to the crystal axes).  See \cite{Korotchenko} for details.

 As noted above, matrix element (\ref{eq20}) and formulae (\ref{eq22}-\ref{eq23},\ref{eq26},\ref{eq27}) are functions of angles $\Theta$ and $\Phi$. Therefore, the probability of radiation of a TWcr-photon has a parametric dependence on these angles. For simplicity, we call this dependence the ``angular distribution''.

\begin{figure}[h]
\centering\noindent
\includegraphics[width=7cm]{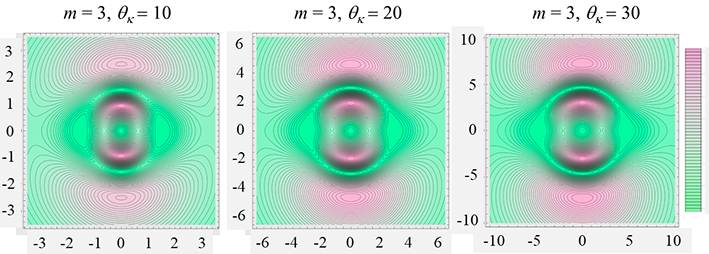}
\includegraphics[width=7cm]{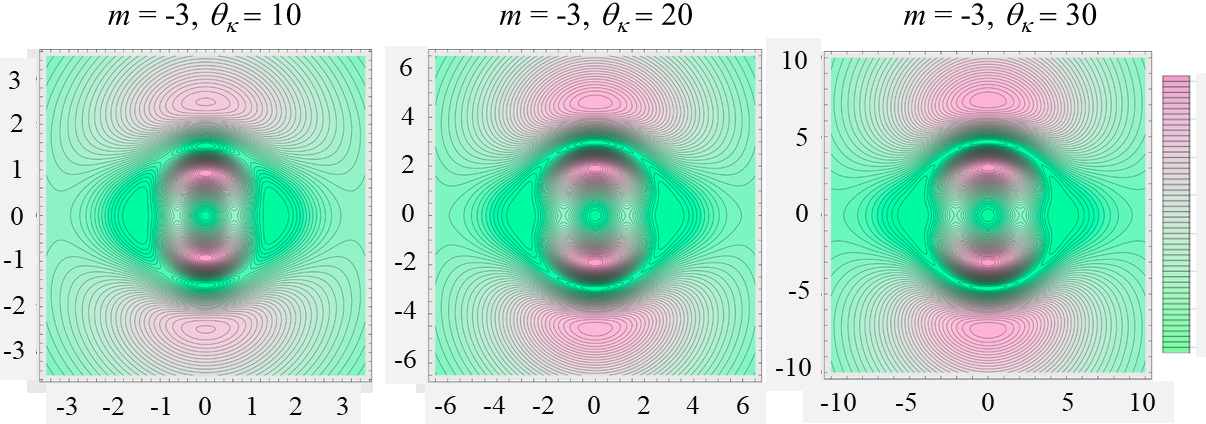}
\caption{Angular distribution of TWcr-photons with TAM $z$-projection $m = \pm 3$ for ``internal'' angles $\theta_\kappa$,  from $\theta_\kappa = 10^{\circ}$ up to $\theta_\kappa = 30^{\circ}$ (the arbitrary units have been used).}\label{fig:3}
\end{figure}

 Figure~\ref{fig:3} shows the calculated angular distributions of TWcr-photons with $z$-projection of total angular momentum (TAM) $m = \pm 3$ and ``internal'' angles $\theta_\kappa$, in the range from from $\theta_\kappa = 10^{\circ}$ to $\theta_\kappa = 30^{\circ}$. Here and below, the calculation was performed for the $10$ MeV electrons channeled along the $\langle 100\rangle$ axes of the Si crystal (as in \cite{Korotchenko}).

 The angular distributions of TWcr-photons with negative and positive TAM practically coincide in accordance with Eqs. (\ref{eq27} - \ref{eq28}).

\begin{figure}[h]
\centering\noindent
\includegraphics[width=7cm]{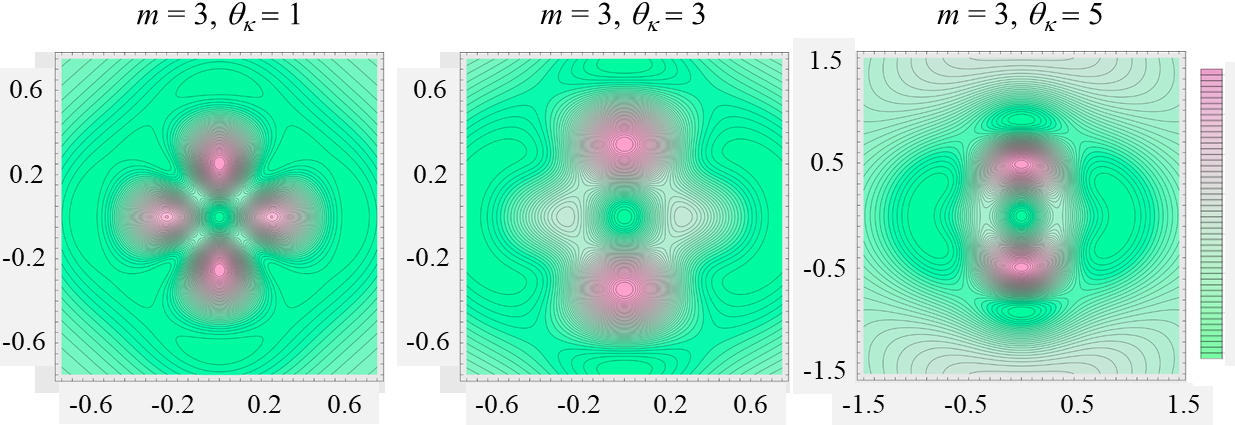}
\includegraphics[width=7cm]{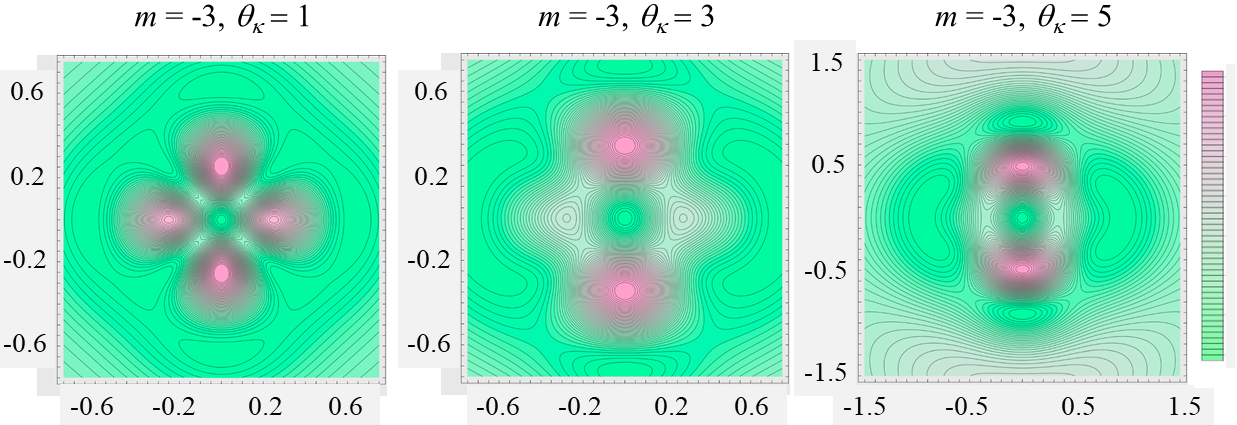}
\caption{Angular distribution of TWcr-photons with TAM $z$-projection $m = \pm 3$ for extremely small angles $\theta_\kappa$,  from $\theta_\kappa = 1^{\circ}$ up to $\theta_\kappa = 5^{\circ}$ (the arbitrary units have been used).}\label{fig:4}
\end{figure}

 The angular distributions of TWcr-photons with TAM $z$-projection $m = \pm 3$, but for extremely small ``internal'' angles $\theta_\kappa$, in the range from $\theta_\kappa = 1^{\circ}$ to $\theta_\kappa = 5^{\circ}$ are presented in Fig.~\ref{fig:4}.

 As noted above, as the angle $\theta$ tends to zero the TWcr radiation turns into CR. This is clearly seen from the distribution, calculated for the ``internal'' angle $\theta_\kappa = 1^{\circ}$. The cylindrical symmetry of the angular distribution is broken and coincides with that for the CR \cite{Korotchenko} at the axial channeling. Figure~\ref{fig:4} illustrates that the shape of the angular distribution of TWcr-photons changes with increasing ``internal'' angle.
\begin{figure}[h]
\centering\noindent
\includegraphics[width=7cm]{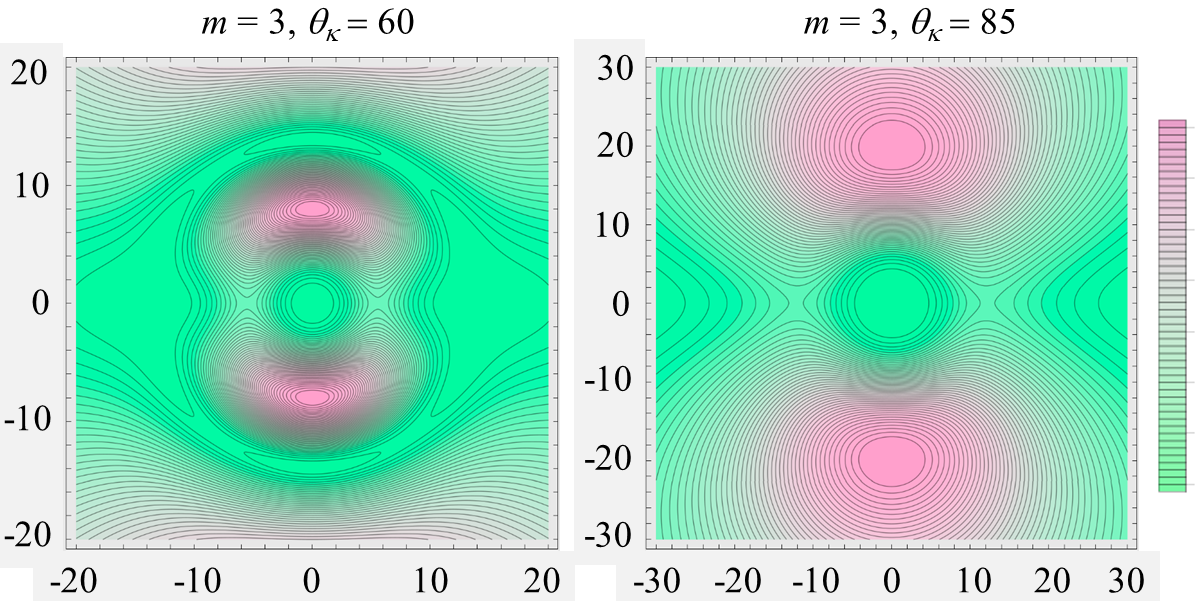}
\includegraphics[width=7cm]{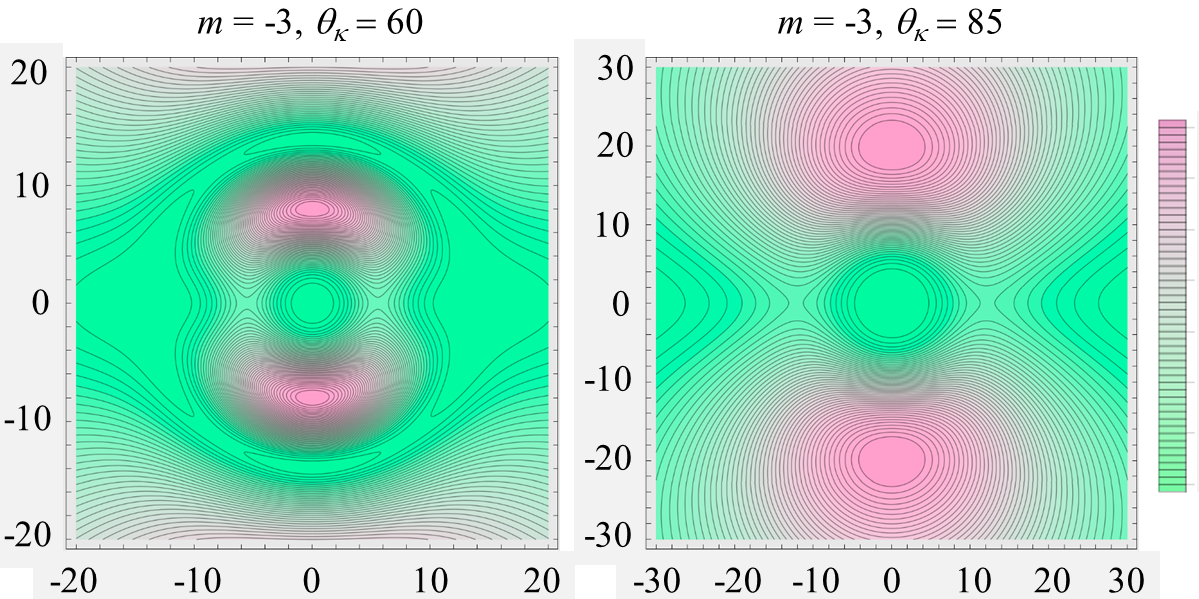}
\caption{Angular distribution of TWcr-photons with TAM $z$-projection $m = \pm 3$ for ``internal'' angles $\theta_\kappa = 60^{\circ}$ and to $\theta_\kappa = 85^{\circ}$ (the arbitrary units have been used).}\label{fig:5}
\end{figure}

 At the small ``internal'' angle $\theta_\kappa$, the angular distributions of TWcr photons with positive and negative TAM values (Fig.\ref{fig:4}) are slightly different (especially for $\theta_\kappa = 1^{\circ}$). These differences become insignificant as the ``internal'' angle increases.
 Figure~\ref{fig:5} shows the results of calculating the angular distributions of TWcr-photons with $m = \pm 3$, but for large ``internal'' angles $\theta_\kappa = 60^{\circ}$ and $\theta_\kappa = 85^{\circ}$.
 The angular distribution's cylindrical symmetry has been broken, as can be seen. The change in symmetry begins with the  ``internal'' angle $\theta_\kappa = 60^{\circ}$. For the ``internal'' angle $\theta_\kappa = 85^{\circ}$, there are two radiation peaks located  near points on the $Y$-axis.
\begin{figure}[h]
\centering\noindent
\includegraphics[width=7cm]{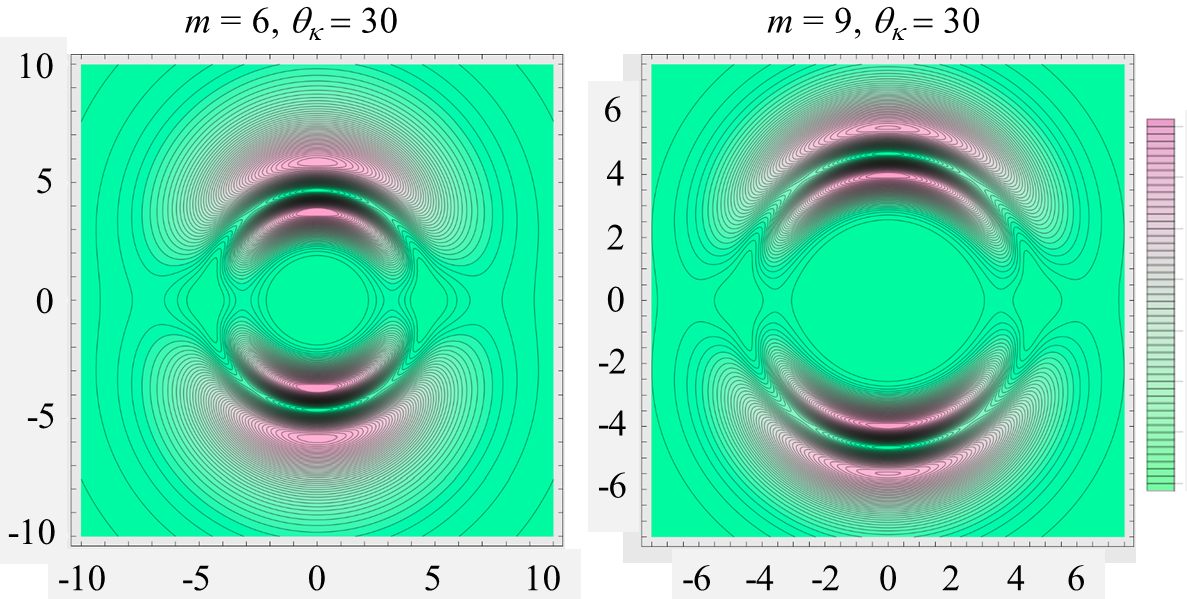}
\caption{Angular distribution of TWcr-photons with TAM $z$-projection $m = 6$ and $m = 9$ for ``internal'' angles $\theta_\kappa = 30^{\circ}$ (the arbitrary units have been used).}\label{fig:6}
\end{figure}

 Finally, Fig.~\ref{fig:6} shows the angular distributions of TWcr-photons for two different TAM $z$-projections $m = 6$ and $m = 9$ for the ``internal'' angle $\theta_\kappa = 30^{\circ}$. The more TAM of TWcr-photons the more pronounced the cylindrical symmetry (compare with Fig.~\ref{fig:3}).

\section{Conclusion}\label{sec:6}

 Within the framework of QED, we have developed for the first time the theory of emission of twisted photons by an axially channeled electron at arbitrary angle with respect to electron longitudinal momentum. The TWcr angular distribution is compared with the angular distribution of CR.

In summary, we note:
\begin{itemize}
  \item The wave function of the TWcr-photon depends on the angles $\Theta$ and $\Phi$. These angles are additional parameters to the usual wave function variables. As a result, the matrix element and the probability of TWcr-photon emission depend parametrically on these. For simplicity, we call this dependence the ``angular distribution''.
  \item The angular distributions of TWcr, unlike CR, have cylindrical symmetry. The more TAM of TWcr-photons the more pronounced the cylindrical symmetry. The angular distribution of TWcr-photons with negative and positive TAM values is almost identical.
  \item The cylindrical symmetry of the angular distribution is broken at a large ``internal'' angles $\theta_\kappa$. For example, at $\theta_\kappa = 85^{\circ}$ there are two emission peaks located near points on the Y axis. However, the TWcr angular distribution differs greatly from that of CR.
  \item As the angle $\theta_\kappa$ tends to zero, the TWcr-photon becomes an ordinary photon, and the probability of the TWcr-photon emission must coincide with the probability of CR emission. Calculations show that in this case the angular distribution of the TWcr-photon is similar to the CR distribution. At a small internal angle, there is a slight difference in angular distributions between TWcr-photons with positive and negative TAM values.
\end{itemize}

 Despite a large number of experimental works on CR, twisted photons have not been observed in channeling conditions. The main reason for this, on the one hand, is the problem of registration twisted photons, and, on the other hand, the lack of experimental works on TWcr investigation. We believe that the difference in angular distributions can be used in experiments to distinguish TWcr from CR background.

 The emission of electromagnetic radiation with high TAM electrons channeled in crystals has already been discussed, but within the framework of classical electrodynamics \cite{SAbdrash}, \cite{Kazin}.
 In the literature, there has already been a discussion about the emission of electromagnetic radiation with high TAM from electrons channeled in crystals, but within the framework classical electrodynamics \cite{SAbdrash}, \cite{Kazin}.

% According to \cite{SAbdrash}, \cite{Kazin}, the planar case is the most attractive from an experimental standpoint. The plane case will be considered in our next work.

\bibliographystyle{plain}

\end{document}